\documentclass[12pt,preprint]{emulateapj}
\shorttitle{Evolving Winds in GRO J1655-40}
\shortauthors{Neilsen et al.}

\begin{document}

\title{A Hybrid Magnetically/Thermally-Driven  Wind in the Black Hole GRO
  J1655--40?} 

\author{Joseph Neilsen\altaffilmark{1}, Jeroen Homan\altaffilmark{1}}
\altaffiltext{1}{MIT Kavli Institute for Astrophysics and Space Research,
  Cambridge, MA 02139; jneilsen@space.mit.edu}

\begin{abstract}
During its 2005 outburst, GRO J1655--40 was observed twice with the
\textit{Chandra} High Energy Transmission Grating Spectrometer; the
second observation revealed a spectrum rich with ionized absorption
lines from elements ranging from O to Ni
(\citealt{M06a,M08,Kallman09}), indicative of an outflow too dense and
too ionized to be driven by radiation or thermal pressure. To date,
this spectrum is the only definitive evidence of an ionized wind
driven off the accretion disk by magnetic processes in a black hole
X-ray binary. Here we present our detailed spectral analysis of the
first \textit{Chandra} observation, nearly three weeks earlier, in
which the only signature of the wind is the Fe\,{\sc xxvi} absorption
line. Comparing the broadband X-ray spectra via photoionization
models, we argue that the differences in the \textit{Chandra} spectra
cannot possibly be explained by the changes in the ionizing spectrum,
which implies that the properties of the wind cannot be constant
throughout the outburst. We explore physical scenarios for the 
changes in the wind, which we suggest may begin as a hybrid
MHD/thermal wind, but evolves over the course of weeks into two  
distinct outflows with different properties. We discuss the
implications of our results for the links between the state of the
accretion flow and the presence of transient disk winds. 

\end{abstract}
                 
\keywords{accretion, accretion disks --- black hole physics --- X-rays:
  individual (GRO J1655--40) ---  X-rays: binaries --- stars: winds,
  outflows}

\section{INTRODUCTION}
\label{sec:intro}
The last two decades have seen the discovery of a multitude of
highly-ionized absorbers and winds in the X-ray spectra of
accreting black hole and neutron star X-ray binaries
(e.g.\ \citealt{Ebisawa97a,Kotani97a,BS00,K00,Kotani2000a,L02,
  Schulz02,U04,M04a,M06a,M06b,M08,NL09,U09,Blum10,ReynoldsM10,
  Miller11,N11a,N12a,King11}). In many of these cases, although the
kinetic power in these outflows is orders of magnitude below the
accretion power, the mass loss rate in the wind can be much greater
than the accretion rate at the inner edge of the disk
(e.g.\ \citealt{N11a}). The ubiquity of disk winds in X-ray binaries
and their extreme mass loss rates suggests that these outflows play a
crucial role in the physics of accretion. 

Presently, accretion disk winds are  understood to be driven
by some combination of radiation, thermal, and magnetic pressure,
which operate in different regimes of density and ionization
(\citealt{B83,Proga2000,PK02,M06a}). Yet determining which of these
processes dominates the physics of an individual outflow typically
presents a significant challenge. Most often, indirect density
estimates must come from geometrical arguments that constrain the
distance $R$ of the wind from the black hole
(e.g.\ \citealt{L02,N11a,N12a}). In these cases, the density $n$ can
be found using the observed ionization parameter $\xi$ and luminosity
$L$ (\citealt{Tarter69}): 
\begin{equation}\label{eq:xi}
\xi=\frac{L}{nR^2}.
\end{equation}

\defcitealias{M06a}{M06}
\defcitealias{M08}{M08}
\defcitealias{Kallman09}{K09}
In rare cases it may be possible to measure the density directly from
density-sensitive atomic lines, a technique that was most notably
applied in the discovery of a magnetically-driven wind in GRO
J1655--40 (\citealt{M06a}, hereafter \citetalias{M06a}). When
\textit{Chandra} observed this microquasar with the High Energy
Transmission Grating Spectrometer (HETGS; \citealt{C05}) in the middle
of its 2005 outburst, the high-resolution X-ray spectrum (ObsID 5461)
revealed a rich series of absorption lines from a dense
($n\gtrsim10^{14}$ cm$^{-3}$), highly-ionized ($\xi\sim10^4$)
wind. Detailed follow-up spectral analysis and theoretical studies
(\citealt{M08}, hereafter \citetalias{M08};
\citealt{Kallman09}, hereafter \citetalias{Kallman09};
\citealt{Luketic10}) have repeatedly confirmed the original 
conclusion of \citetalias{M06a}: this outflow was driven primarily via
magnetic processes, rather than radiation or thermal pressure. 

\begin{figure}
\centerline{\includegraphics[width=3.3 in]{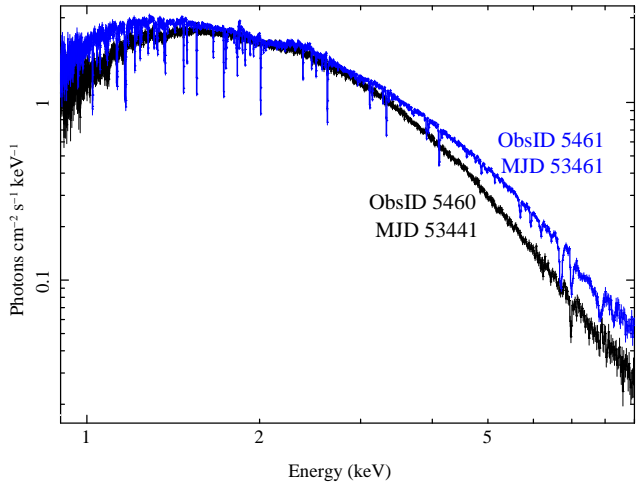}}
\caption{\textit{Chandra} HETGS spectra of GRO J1655--40 during its
  2005 outburst. Shown in black is the spectrum of the first
  observation, when the broadband X-ray continuum was hard and very
  few absorption lines were present; shown in blue is the spectrum of
  the second observation, which was executed 20 days later and
  recorded a rich series of lines from the accretion disk wind. In
  this paper, we address the physics driving these significant changes
  in the disk wind. \label{fig:hetgs}}
\end{figure}

However, a HETGS observation (ObsID 5460) of the same outburst made 20
days earlier reveals an outflow that is empirically quite
different (see Figure \ref{fig:hetgs}). Instead of the host of lines,
this observation merely contains an Fe\,{\sc xxvi} absorption line
noted in passing by \citetalias{M08}. In this paper, we ask: where did the
other lines go? During the first observation, are we simply observing
a much more ionized (and thus transparent) version of the wind
studied by \citetalias{M06a}, \citetalias{M08}, and
\citetalias{Kallman09}? Or do these two observations actually probe a
transient wind that evolves physically and geometrically throughout
the black hole outburst? 

We describe the observations and data analysis in Section
\ref{sec:obs}. In Section \ref{sec:spectra}, we present the
\textit{Chandra} spectra and use broadband X-ray spectra from the
\textit{Rossi X-ray Timing Explorer}'s (\citealt{J96}) Proportional
Counter Array (\textit{RXTE} PCA) to study the ionizing continuum. We
also build photoionization models to demonstrate how the wind responds
to changes in the incident X-ray spectrum. In Section
\ref{sec:discuss}, we discuss the implications for the evolution 
of wind physics during the 2005 outburst. We conclude in Section
\ref{sec:concl}. 
%\clearpage
%\clearpage
 
\section{OBSERVATIONS AND DATA REDUCTION}
\label{sec:obs}
GRO J1655--40 was observed with the \textit{Chandra} HETGS on 2005
March 12 (20:42:53 UT; ObsID 5460) and 2005 April 1 (12:40:41 UT;
ObsID 5461) for 34.3 and 62.15 ks, respectively. The data were taken
in Graded Continuous Clocking mode, which has a time resolution of
2.85 ms and is designed to limit both photon pileup and telemetry
saturation.

We reduce and barycenter-correct the \textit{Chandra} data with
standard tools from the {\sc ciao} analysis suit, version 4.3. We use
the order-sorting routine to remove the readout streak on the chip S4,
since the dedicated tool (\textit{destreak}) can also remove source
counts when the count rate is high. After reprocessing, we extract
High-Energy Grating (HEG) spectra and make grating response files. 

During the \textit{Chandra} observations, \textit{RXTE} made pointed
observations of GRO J1655--40 with good exposure times of 1.82 and
12.08 ks, respectively. In this paper, we analyze data from the PCA,
which covers the 2--60 keV band, in order to explore the role of the
photoionizing continuum in driving the evolution of the wind. For this
spectral analysis, we extract Standard-2 129-channel spectra, and
restrict our attention to the 3--45 keV spectrum from PCU2. We assume
0.6\% systematic errors in each energy bin. All our spectral fitting
is done in ISIS (\citealt{HD00,Houck02}). Following \citetalias{M08},
we assume a neutral hydrogen column density $N_{\rm
  H}=7.4\times10^{21}$ cm$^{-2}$ and a distance of 3.2 kpc
(\citealt{Orosz97}). 
\section{X-ray Spectroscopy}
\label{sec:spectra}
The \textit{Chandra} HEG spectra are shown in Figure \ref{fig:hetgs},
where it is abundantly clear that there are very significant
differences between the absorber during the first (black) and second
(blue) observations. In this section, we characterize the
high-resolution and broadband spectra in an effort to determine
whether photoionization alone can explain explain the differences in
the lines. Our method is as follows: first we model the HEG/PCA
spectra, then use {\sc xstar} to calculate the expected atomic level
populations for a gas slab illuminated by our observed ionizing
continuum. Finally, we compare the predicted and observed absorption
spectra. We focus on the first observation, which has not been studied 
in detail. 
 %\clearpage
\begin{figure}
\centerline{\includegraphics[width=3.3 in]{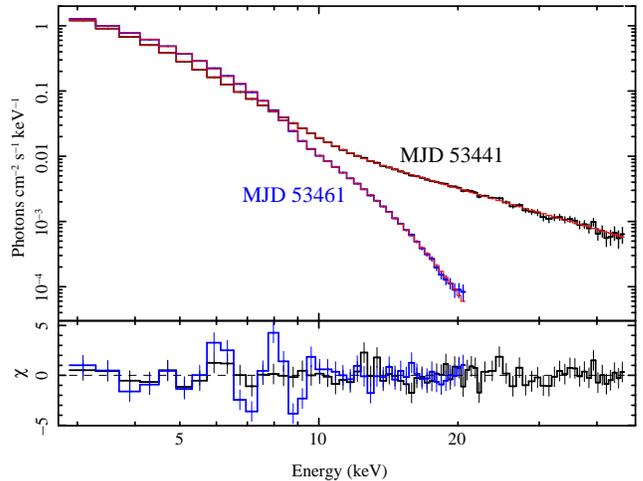}}
\caption{\textit{RXTE} PCA spectra of GRO J1655--40 during its
  2005 outburst. Shown in black is the spectrum of the first
  observation, when the broadband X-ray continuum was hard and very
  few absorption lines were present; shown in blue is the spectrum of
  the second observation, which occurred 20 days later and exhibits a
  significantly softer X-ray continuum. In Section \ref{sec:spectra}
  we argue that the changes in photoionizing flux are insufficient
  to explain the changes in the lines (Figure
  \ref{fig:hetgs}).\label{fig:pca}}  
\end{figure}
%\clearpage
\noindent
\subsection{Preliminary Modeling}
\label{sec:hetgspca}
To start, we fit the HEG spectra phenomenologically with an absorbed
polynomial spectrum and multiplicative Gaussian absorption lines
(multiplicative absorption models are accurate even if the lines are
moderately optically thick; \citealt{N12a}). The primary goal of this
phenomenological modeling is not to catalog the lines or investigate
their properties in detail (which has been done several times for the
second observation, e.g.\ \citetalias{M08,Kallman09}), but to incorporate
the ionized lines into models of the broadband X-ray spectrum as seen
by \textit{RXTE}. Thus we devote relatively little space in this 
section to a detailed discussion of the lines themselves. 

However, a direct comparison of the Fe\,{\sc xxvi} Ly$\alpha$ lines in
the two observations is informative. In both spectra, we achieve tighter
constraints on the parameters of these lines by fitting the Ly$\alpha$
and Ly$\beta$ lines simultaneously (i.e.\ with the same ion column
density, velocity, and line width), since the optical depth in a line
from any given ion is proportional to the line oscillator strength
times the ion column density divided by the line width. Thus, even
though Ly$\beta$ is not observed during observation 1, its absence
places an upper limit on the column density of Fe\,{\sc xxvi}. In the
first observation, we find a column density $N_{\rm Fe\,{\sc xxvi}}
=6_{-1}^{+2}\times10^{17}$ cm$^{-2}$ and a blueshift of $v=-470\pm230$
km s$^{-1}$ (the errors are 90\% confidence limits). The line width is
$\sigma=13_{-7}^{+8}$ eV, and the equivalent width is $W_{0}\approx24$
eV. In the second observation, the Fe\,{\sc xxvi} has a similar column
density, $N_{\rm Fe\, {\sc xxvi}} =(7.6\pm0.9)\times10^{17}$ cm$^{-2}$, but 
a higher velocity, width, and equivalent width ($v=-1350^{+170}_{-160}$ 
km s$^{-1},~\sigma=21\pm4$ eV, and $W_{0}\approx31$ eV). In the second 
observation, at least 33 additional strong features are present (in
the 2.5-9 keV range) that are not visible in the first dataset.  

In short, we see that the Fe\,{\sc xxvi} column densities are similar,
but all the other ion column densities increase dramatically from the
first observation to the second, 20 days later. The simplest
explanation for this fact is that the wind is more ionized during the
earlier pointing. To test this hypothesis in more detail, we
turn our attention to  the broadband PCA spectra, which are shown in
Figure \ref{fig:pca}. From this figure, the changes in the ionizing
continua between the two observations are immediately apparent:
during the first observation, there were many more photons capable of 
ionizing hydrogen- and helium-like iron (i.e.\ with $E\gtrsim9$
keV). On the other hand, other elements have very small
photoionization cross-sections for $E\gtrsim9$ keV. So if the geometry
or bulk properties of the wind were constant in time, we would expect
there to be relatively little difference between the two observations
in terms of absorption lines from ions other than iron. Since we
observe highly-significant changes in lines other than iron, we can already conclude that
excess ionizing luminosity alone is definitely not responsible for the
absence of lines in the first observation.  

Our best fit model for the broadband continuum in the first
observation ({\tt tbabs*(simpl $\otimes$ezdiskbb
+egauss)*nmgauss*edges}) consists of an absorbed ({\tt tbabs};
\citealt*{Wilms00}) hot disk ({\tt ezdiskbb}; \citealt{Zimmerman05})
convolved through a scattering kernel  ({\tt simpl};
\citealt{Steiner09a}), a Gaussian emission line at 6.4 keV, along with
the Fe\,{\sc xxvi} absorption line ({\tt nmgauss}; \citealt{N12a})
detected in the HEG spectrum. {\tt simpl} takes a seed spectrum and
scatters a fraction $f_{\rm SC}$ of the source photons into a power
law, approximating the high-temperature, low optical depth regime of 
Comptonization. Following \citet{Saito06}, we allow for possible
ionized absorption edges at 7.7, 8.83, 9.28, and 10.76 keV, but none
of these are detected at 90\% confidence. Assuming a disk inclination
of $67^{\circ}$ (\citealt{Orosz97}) and a color correction factor of
1.7, the disk normalization $N=390\pm20$ implies an inner disk radius
$R_{\rm in}=29.1_{-0.6}^{+0.7}$ km, or roughly 2 gravitational radii,
and the disk temperature is $T=0.973\pm0.009$ keV. We find that
$(5.6\pm0.3)$\% of the disk photons are scattered into the power law,
which has photon index $\Gamma=2.11\pm0.04.$ The implied unabsorbed
luminosity ($10^{-3}$--$10^3$ keV) is $L=6.4\times10^{37}$ ergs s$^{-1},$
roughly 7\% of the Eddington limit. For the second observation, we
include a {\tt highecut} term to account for curvature in the
spectrum; we can also replace {\tt simpl$\otimes$ezdiskbb} with {\tt
  ezdiskbb+nthcomp} (\citealt{Zdziarski96,Zycki99}). For our purposes
here, only the luminosity of the second observation is relevant:
$L\approx5.7\times10^{37}$ ergs s$^{-1}.$ 
\subsection{Photoionization Modeling}
\label{sec:warmabs}
The immediate goal of our analysis is to determine why there are so
few lines in the first observation. Are the changes in the luminosity
and spectral shape enough to wipe out the lines? In Section
\ref{sec:hetgspca} we argued that the answer is no, and that changes
in the wind geometry or density are required. In this section, we make
this argument quantitative with photoionization models.  

The method itself is straightforward. We use {\sc xstar} to calculate
atomic data for an illuminated slab of gas. We define this slab of gas
to match the properties of the wind seen in the \textit{second}
observation, as modeled by \citetalias{Kallman09}, and we illuminate
it with the broadband X-ray spectrum of the \textit{first}
observation. In this way, we can test whether the same wind, ionized
by a harder X-ray spectrum, should have been visible in the first
dataset. For reference, we base our models on Model 6 of
\citetalias{Kallman09}\footnote{We compare our results to
\citet{Kallman09} rather than the original models of \citet{M06a,M08}
because the newer models (at a higher density) provide a superior fit
to the spectrum.}: a density of $n=10^{15}$ cm$^{-3}$, an 
equivalent column density $N_{\rm H}=10^{24}$ cm$^{-2}$, a 37\%
partial covering fraction, a blueshift of $v=375$ km s$^{-1}$, and a 
turbulent line width of $v_{\rm turb}=200$ km s$^{-1}.$ We increase
the ionization parameter from $\xi=10^4$ to $\xi=10^{4.05}$ to account
for the higher luminosity during the first observation. Finally, we
use the elemental abundances reported in Figure 21 of
\citetalias{Kallman09}. 

\begin{figure}
\centerline{\includegraphics[width=3.3 in]{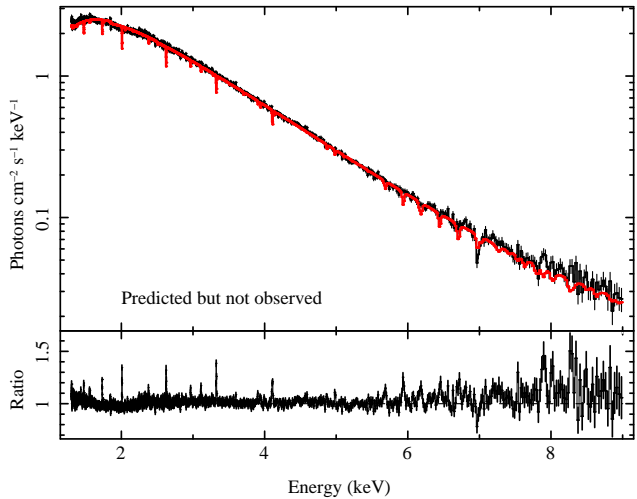}}
\caption{\textit{Chandra} HETGS spectrum of the first observation of
  GRO J1655--40 (black) with model (red) based on the wind described
  by \citet{Kallman09}. If the wind were the same during both
  observations, a number of strong lines would have been seen during
  the first observation, despite its harder spectrum. That these lines
  are not observed indicates that the wind evolved significantly
  throughout the outburst. \label{fig:predict}}
\end{figure}

We incorporate the resulting atomic data into new models of the HETGS
observations using the analytic fit function {\tt warmabs}. Using
parameters equivalent to those measured by \citetalias{Kallman09}, but
with an ionization balance appropriate for the first dataset, we find
that there are a number of lines that would have been detected at high
significance if the bulk properties of the wind were the same as those
measured by \citetalias{Kallman09} (see Figure
\ref{fig:predict}). Three of the strongest features 
predicted by this test model but not observed in our data are 
Si\,{\sc xiv} at 2.0, S\,{\sc xv} at 2.6 keV, and Ar\,{\sc xviii} at
3.2 keV. To give an idea of how strong the features are, we note that
without introducing any additional curvature into the spectrum (only
lines), this test model increases $\chi^2$ by over 1800 relative to a
model with no wind at all. This result confirms our earlier
conclusion: changes in the spectral shape and luminosity cannot be
solely responsible for the absence of spectral lines in the first
observation. 

\section{DISCUSSION}
\label{sec:discuss}
The results of our photoionization models in Section \ref{sec:warmabs}
are unequivocal: if the bulk properties of the wind were constant in
time during the 2005 outburst of GRO J1655--40, a large number of
strong lines should have been visible in the first dataset. A constant
wind, simply over-ionized in the hard state, is ruled
out. Furthermore, since the observations occur at orbital phases 0.5 and
0.0 (where the disk would be eclipsed at 0.75; \citealt{Orosz97}),
orbital or companion star effects seem unlikely to be important. These 
results have broad implications for the role of winds in outburst
(e.g.\ in quenching jets; \citealt{NL09}). In this period of three
weeks, as the source moved out of the hard, jet-producing state into a
spectrally-soft state, the structure, density, or geometry of the
accretion disk wind must have changed significantly. But what else can
be said about the details of this evolving wind, given that so few
lines are visible in the first dataset? 

Most of what we know for certain comes from observation 2, which shows
not only clear indicators of an MHD wind, but also compelling evidence
for a second wind component: \citet{Kallman09} found that while the
typical blueshift of absorption lines in the spectrum is $\sim400$ km
s$^{-1}$, the strong lines of Fe\,{\sc xxvi} and Ni\,{\sc xxvii} are
blueshifted by $\sim1300$ km s$^{-1}$. See their Figure 15 for a
remarkable demonstration of this result, which they interpreted %\clearpage
%\clearpage
\noindent as indicative of a second, extremely highly ionized
component. In fact, similar (although less extreme) differences in
line dynamics have also been seen in other X-ray binaries. For
example, studying an equally rich absorption line spectrum in GRS
1915+105, \citet{U09} found that while most lines were narrow and
moderately blueshifted ($\sigma\sim70$ km s$^{-1}$, $v\sim150$  km
s$^{-1}$), the Fe\,{\sc xxvi}  line was broader and faster
($\sigma\sim200$ km s$^{-1}$, $v\sim500$ km s$^{-1}$). Thus we are
convinced that there is a two-component outflow in the second
observation of GRO J1655--40. Since the magnetically-driven wind is a
dense outflow at small radii, we suppose that the second component is
a relatively low-density ($n\lesssim10^{12}$ cm$^{-3}$) outflow at
large radii ($R\gtrsim10^{11}$ cm), i.e.\ as produced by thermal
driving. 

But how does this color our understanding of observation 1? Any
explanation must hinge on the Fe\,{\sc xxvi} line, which was present
in both pointings. We focus on the bulk properties of the wind
(i.e.\ density, location, and so on) in the hopes of determining the
contributions of thermal and magnetic processes to the outflows in GRO
J1655--40 (radiation pressure is ruled out for the same reason as in
observation 2: it is ineffective at such high ionization
parameters). So in short, it comes down to whether we believe the iron 
absorption line in the first spectrum should be associated with the
slow, dense MHD component or the faster, more ionized thermal
component in the second spectrum. Thus the interpretation of this line
can be roughly separated into three scenarios, in which the wind in
observation 1 is a purely thermal wind (Section \ref{sec:thermal}), a
magnetically-driven wind (already discussed in Section
\ref{sec:warmabs}), or a hybrid thermal/MHD wind (Section
\ref{sec:evolve}). 

While there is a very large body of literature on the theoretical
aspects of the origin and properties of accretion disk winds
(e.g.\ \citealt{B83,Shields86,Woods96,Proga2000,PK02,Luketic10}),
there is relatively little information about which wind processes are
expected to be most important at any given phase of a black hole
outburst (cf.\ the unified model of disk-jet coupling in outburst;
\citealt{FBG04}). For example, are there links between the accretion
state and the relative importance of radiatively-, thermally-, and
magnetically-driven winds? Observations (\citealt{N12a,Ponti12})
suggest that varying illumination and shadowing of the outer disk may
be important in producing transient winds, but it is unclear how the
density, launch radius, and geometry of these winds (MHD or thermal)
should evolve over time. In the following subsections,  we explore the
implications of a variety of simple assumptions about the behavior of
the wind in GRO J1655--40. By no means do these constitute an
exhaustive list, but they should be representative of the
possibilities.  
\begin{deluxetable}{ccccccc}
\tabletypesize{\scriptsize}
\tablecaption{Photoionization Models of GRO J1655--40
\label{tbl:mod}}
\tablewidth{0pt}
\tablehead{
\colhead{$\log$}  &
\colhead{$\log$}  & 
\colhead{$\log$}  &
\colhead{$\log$}  & 
\colhead{$\log$}  &
\colhead{$\log$}  & 
\colhead{$\log$} \\
\colhead{$n_{\tt w}$}  &
\colhead{$N_{\rm H}$}  &
\colhead{$\xi$}  &
\colhead{$n_{\Delta R}$}  &
\colhead{$R_{\Delta R}$} & 
\colhead{$n_{\Delta R/R}$}  &
\colhead{$R_{\Delta R/R}$} 
}
\startdata
11 & $23.2_{-0.2}^{+0.2}$ & $4.5_{-0.1}^{+0.2}$ & 14.2 & 9.5 & 13.9 & 9.7 \\
12 & $22.7_{-0.2}^{+0.2}$ & $4.3_{-0.1}^{+0.2}$ & 13.7 & 9.9 & 12.8 & 10.4 \\
13 & $22.6_{-0.2}^{+0.1}$ & $3.9_{-0.1}^{+0.2}$ & 13.6 & 10.2 & 12.0 & 11.0 \\
14 & $22.8_{-0.2}^{+0.1}$ & $4.0_{-0.1}^{+0.2}$ & 13.8 & 10.0 & 12.5 & 10.7 \\
\vspace{1mm}
15 & $23.0_{-0.2}^{+0.2}$ & $4.2_{-0.1}^{+0.2}$ & 14.0 & 9.8 & 13.2 & ~10.2 
\enddata
\tablecomments{$n$ is the electron density input for {\sc xstar} in
  cm$^{-3};$ $N_{\rm H}$ and $\xi$ are the best-fit column density and
  ionization parameter for observation 1 in cm$^{-2}$ and ergs cm
  s$^{-1}$; $n_{\Delta R}$ and $R_{\Delta R}$ are the density
  (cm$^{-3}$) and radius (cm) implied by the corresponding value of 
  $N_{\rm H}$ if $\Delta R$ is fixed for the wind (Section
  \ref{sec:dr}); $n_{\Delta R/R}$ and $R_{\Delta R/R}$ are the density
  and radius implied by $N_{\rm H}$ if $\Delta R/R$ is fixed (Section
  \ref{sec:drr}).}  
\end{deluxetable}

\subsection{Case 1: A Thermal Wind in Observation 1}
\label{sec:thermal}
\begin{figure}
\centerline{\includegraphics[width=3.3 in]{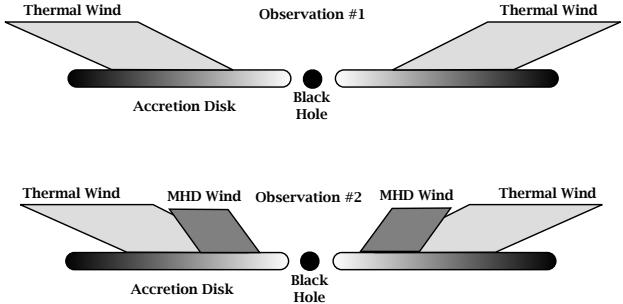}}
\caption{Cartoon of the scenario presented in Section
  \ref{sec:thermal}, in which the wind in observation 1 is
  thermally-driven (top), while observation 2 (bottom) features a
  two-component wind, i.e.\ MHD and thermal. For the winds, darker
  shading implies higher density, and the different orientations of
  the two components reflects our uncertainty about the vertical
  component of the wind velocity. See text for
  details.\label{fig:thm}} 
\end{figure}

Here, in light of the fact that Fe\,{\sc xxvi} is present
in both observations of GRO J1655--40, and may arise in a thermal wind
in observation 2, we consider the possibility that this line also
arises in a thermal wind during observation 1. In other words, we
suppose that the thermal wind is persistent, while the MHD outflow is
transient. This scenario is illustrated in Figure
\ref{fig:thm}. Without decent estimates of the density or ionization
parameter of the thermally-driven wind, it is very difficult to
estimate its location. Thermally-driven winds can be launched anywhere
outside 0.1 Compton radii ($R_{\rm  C}=9.8\times 10^{17}\,M_{\rm BH}\,
T_{\rm IC}^{-1}$ cm; \citealt{B83,Woods96,Rahoui10}), where $M_{\rm BH}$
is the black hole mass in solar masses and the Compton temperature in
Kelvins is given by:
\begin{equation}\label{eq:tic}
T_{\rm IC}=\frac{1}{4k_{\rm B}}\frac{\int_{0}^{\infty}h\nu
  L_{\nu}d\nu}{\int_{0}^{\infty}L_{\nu}d\nu},
\end{equation} where $h$ is Planck's constant, $\nu$ is the frequency,
and $L_{\nu}$ is the monochromatic luminosity. For our PCA spectra of
observations 1 and 2, we find $T_{\rm IC}\sim8.6\times10^{7}$ K and
$8.4\times10^{6}$ K, respectively, which imply Compton radii $R_{\rm
  C}\sim8\times10^{10}$ cm and $8\times10^{11}$ cm. The Compton
temperature of the first observation is much higher because the
spectrum is much harder; accordingly, Compton heating is more
efficient and a thermal wind can arise at smaller radii. Importantly,
in both cases, $0.1R_{\rm C}$ is inside the disk, so that a
Compton-heated wind is plausible at this radius.   

The wind speed provides a different perspective. If we suppose that 
the wind speed is comparable to the local escape speed in the disk,
then the thermal wind  originates at $10^{12}$ cm and $10^{11}$ cm
in the first and second observations, respectively. It is puzzling
that the implied distance between the wind and the black hole in 
observation 1 is comparable to the binary separation. However, the
wind velocity likely has a significant component perpendicular to the
line of sight, so that the Doppler shift represents only a fraction of
the true wind speed. But even if this is the case, the acceleration of
the wind still requires some explanation. Why does the wind accelerate
as the spectrum softens and the luminosity decreases? One possibility
that has gained some ground recently (\citealt{N12a,Ponti12}) is that
during harder states, the geometrically-thick inner disk
(e.g.\ \citealt{Meier01}) shadows some of the outer disk. This
shadowing interpretation is potentially testable via optical/infrared
measures of the irradiation of the companion star or outer disk (a
paper on multiwavelength observations of this outburst is
forthcoming). In this framework, as the scale height of the inner disk
decreases during the state transition in GRO J1655--40, progressively
smaller radii in the disk are irradiated, leading to a faster
wind. A different irradiation geometry could also influence the
direction of the wind velocity, e.g.\ driving it more into the line of
sight in observation 2.
%\clearpage
%\clearpage

In summary, taken at face value, the estimates of $R$ from the Compton
radius and the escape velocity are not fully consistent. However, it
may be possible to reconcile them by considering the vertical
component of the wind velocity (which may be significant). Without a
clear and unambiguous way to estimate the location or the density of
this thermal wind, there is little more we can say about it.
Nevertheless, we feel our estimates here are still merited, as 
there is compelling evidence for a two-component wind 20 days later  
(\citealt{Kallman09}).

\subsection{Case 2: An Evolving Hybrid Wind}
\label{sec:evolve}
\begin{figure}
\centerline{\includegraphics[width=3.3 in]{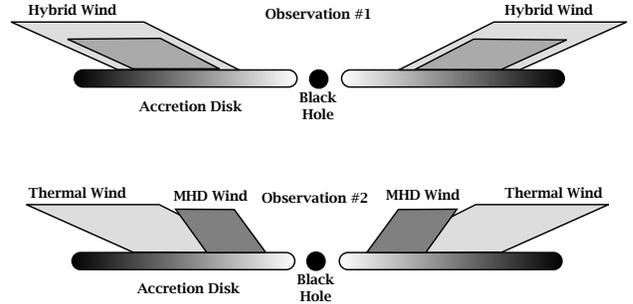}}
\caption{Cartoon of the scenario presented in Section
  \ref{sec:evolve}, in which observation 1 (top) probes a hybrid wind
  driven by a combination of MHD and Compton heating. In this
  interpretation, the hybrid wind evolves over the course of weeks
  into two distinct outflows: a thermal wind at large radii and a MHD
  wind at small radii (bottom). See text for details, and see Figure
  \ref{fig:thm} for notes about shading and orientation in these
  cartoons. \label{fig:evlv}}
\end{figure}

On the other hand, given that observation 2 provides the only direct
evidence for a magnetically-driven wind in an X-ray binary, it is also
worth considering the possibility that both wind components (MHD and 
thermal) are present in both observations of GRO J1655--40. In this
scenario, illustrated in Figure \ref{fig:evlv}, a precursor to the MHD
wind makes a weak contribution to the Fe\,{\sc xxvi} line in
observation 1 (which is dominated by the thermal wind), but in the
subsequent weeks evolves into the dense outflow discovered by
\citetalias{M06a}. There is some circumstantial evidence to suggest
that this scenario is plausible.

In particular, observations with \textit{XMM-Newton} on 2005 March 18
and March 27 (i.e.\ six days after \textit{Chandra} observation 1 and
five days before \textit{Chandra} observation 2, respectively) reveal
the presence of an absorber with an ionization parameter that
decreases rapidly over time (\citealt{DiazTrigo07}). Possible
absorption lines were also detected with \textit{Swift}
(\citealt{Brocksopp06}). It is very difficult to make a quantitative
comparison of the winds seen by \textit{Chandra} and
\textit{XMM-Newton} because the Doppler widths of the lines are so
different: $\sigma\gtrsim3000$ km s$^{-1}$ in the \textit{XMM-Newton}
spectra and $\sigma\approx600$ km s$^{-1}$ in the \textit{Chandra}
observations. This difference may influence the apparent column
density and ionization parameters. Nevertheless, Figure 5 of
\citet{DiazTrigo07} indicates that between March 18 and March 27, the
absorber began to resemble the forest of lines found by
\citet{M06a}. In this context, it is also an interesting coincidence
that the blueshift of the Fe\,{\sc xxvi} line in observation 1 is the
same as the blueshift of the slow MHD wind in observation 2.

Thus it is possible that a precursor to this MHD wind was present
during the first \textit{Chandra} observation. To explore this
possibility, we expand on our {\sc xstar} models from Section
\ref{sec:warmabs}. In what follows, we consider a number of scenarios 
for what the properties of the precursor wind could have been during
observation 1, or how they might have evolved during the outburst. To
be clear, when we refer to a quantity as constant, we mean constant in
time.

%\clearpage
%\clearpage

\subsubsection{Wind Density is Constant}
\label{sec:n}
We begin with the assumption that the density of the MHD wind is
constant during the outburst ($n=10^{15}$ cm$^{-3}$), as in the {\sc
xstar} model shown in Figure \ref{fig:predict}. As discussed in
Section \ref{sec:warmabs}, this model is dynamically identical to the
wind in observation 2 (\citealt{Kallman09}), but features an
ionization balance appropriate for observation 1. In Section
\ref{sec:warmabs}, we used this model in a simple test to show that
the wind in observation 1 cannot have the same column density and
ionization parameter as it does in observation 2 (Figure
\ref{fig:predict}). Here, we allow $N_{\rm H}$ and $\xi$ to 
be free parameters and actually fit a {\tt warmabs} model to the 
\textit{Chandra} spectrum of observation 1. Using a $128\times128$
logarithmically-spaced grid covering $\xi=10^3-10^5$ and $N_{\rm
  H}=10^{22}-10^{24},$ we create a confidence map to search for the 
model that best fits this spectrum. The absence of lower-ionization
lines and Fe\,{\sc xxvi} Ly$\beta$ absorption provides upper limits on
the column density of the model and lower limits on the ionization
parameter, respectively. We allow the MHD precursor wind to contribute 
as much as possible to the Fe\,{\sc xxvi} absorption line and assume the
rest comes from the thermal wind. If the column density is too low,
the predicted iron line will be too weak. Furthermore, because the
column density is constrained from above, the ionization parameter
cannot be arbitrarily high and still contribute to the line. Thus the
Fe\,{\sc xxvi} line provides lower limits on the column density and a 
upper bounds on the ionization parameter. For $n=10^{15}$ cm$^{-3}$,
we find an equivalent hydrogen column density $N_{\rm H}=(9\pm4)
\times10^{22}$ cm$^{-2}$ and an ionization parameter
$\log\xi=4.2_{-0.1}^{+0.2};$ the errors are 99\% confidence limits for
a single parameter. This result implies that the column density of the
MHD wind increased by an order of magnitude between the first and
second \textit{Chandra} observations. 

But the column density cannot be the only parameter that changes
between these two observations, since 
\begin{equation}\label{eq:nh}N_{\rm H}=n\,\Delta R,\end{equation}
where $\Delta R$ is the (integrated) radial extent of the absorber. In
general, in order for the column density to change, there must be
variation in the gas density, the extent of the wind, or both. That
is, in order to explain the differences in the \textit{Chandra} lines,
we require a significant change in at least one of these properties. 
In the model described in this section, the density is constant, which
in turn implies from Equation \ref{eq:nh} that $\Delta R$ must have
been an order of magnitude smaller during observation 1 than it was 20
days later. Although this scenario is not impossible, it is unclear
why the wind would have the same density but arise from a much smaller
region of the disk or have a much smaller filling factor (i.e.\ why it
would be so much clumpier). In other words, we cannot rule it out, but
we are hesitant to accept a model in which the column density changes
solely because of changes in $\Delta R$. 
%\clearpage
\begin{figure}
\centerline{\includegraphics[width=3.3 in]{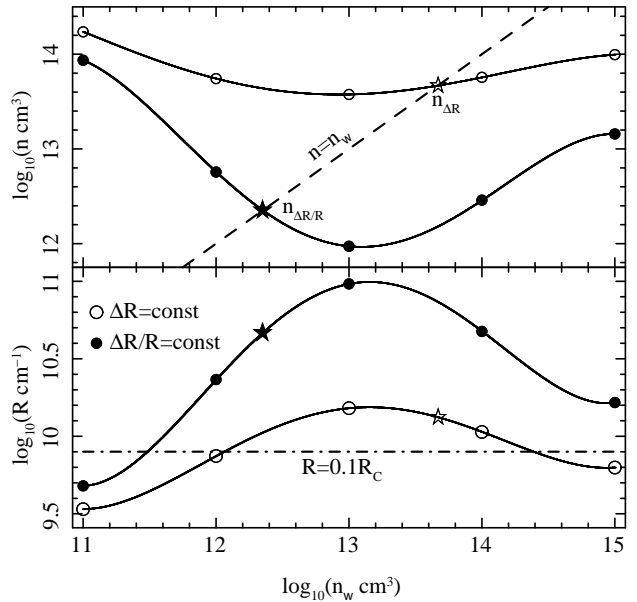}}
\caption{Densities (top) and radii (bottom) implied under various
  assumptions by our grid of {\tt warmabs} models (at densities
  $n_{\tt w}$) for the MHD precursor wind in \textit{Chandra}
  observation 1. The open symbols correspond to the case where $\Delta
  R$ is constant during the outburst (Section \ref{sec:dr}), while the
  filled symbols correspond to the case where $\Delta R/R$ is constant
  (Section \ref{sec:drr}). The solid lines are polynomial
  interpolations of these data points; the dashed line in the top
  panel represents $n=n_{\tt w}.$ The intersection of an interpolation
  and $n=n_{\tt w}$ is a self-consistent solution for the density of
  the wind in observation 1. These self-consistent solutions are shown
  as stars. In the bottom panel, $R=0.1R_{\rm C}$ is shown so that the
  implied radii may be compared to the smallest launch radius for the
  thermal component. See Section \ref{sec:discuss} for
  details. \label{fig:n}}  
\end{figure}
%\clearpage

\subsubsection{$\Delta R$ is Constant}
\label{sec:dr}
At the other extreme is the case where $\Delta R$ is constant, and the
change in the wind column density between the two \textit{Chandra} 
observations is due solely to a change in gas density. In this case,
it is less straightforward to find the density of the wind during
observation 1, because the ionization balance and resulting fits with
{\tt warmabs} are sensitive to the density, meaning that an iterative
process is required. Ideally, to be fully self-consistent, we need the
density $n_{\tt w}$ used to create the {\tt warmabs} model to be the
same as the density $n_{\Delta R}$ implied by Equation \ref{eq:nh},
i.e.:  
\begin{equation}\label{eq:dr}
n_{\tt w}=n_{\Delta R}=\frac{N_{\rm H,1}}{\Delta
  R}=n_{2}\,\frac{N_{\rm H,1}}{N_{\rm H,2}}.  
\end{equation} Subscripts 1 and 2 refer to \textit{Chandra}
observations 1 and 2, respectively. To self-consistently determine the
density for the scenario where $\Delta R$ is constant, we generate an
additional set of {\tt warmabs} models, which we fit to the
\textit{Chandra} spectrum of observation 1. Again, we use the same
covering factor, turbulent line width, and blueshift as
\citet{Kallman09}, but we build models with densities of $\log(n_{\tt
  w}~$cm$^{3})=11,12,13,14,$ and $15.$ For each of these models, we
perform the same grid search and confidence maps described in Section
\ref{sec:n}; the results for $N_{\rm H}$ and $\xi$ are tabulated in
Table \ref{tbl:mod}. Typically, our best fit for any given model
contributes only $\sim3-4$ eV to the equivalent width of the Fe\,{\sc
xxvi} absorption line; the rest we attribute to the thermal component.

As in the preceding subsection, all of these models require that the
column density of the MHD wind in observation 1 be more than an order
of magnitude lower than it was in observation 2. Since we are assuming
here that $\Delta R$ is constant, the density in observation 1 must be
lower by the same factor. The implied densities $n_{\Delta R}$ are
calculated from Equation \ref{eq:dr}, listed in Table \ref{tbl:mod},
and plotted as open circles in the top panel of Figure
\ref{fig:n}; the bottom panel of this figure shows the corresponding
radii from Equation \ref{eq:xi}. Interpolating over our ($n_{\tt
  w},~n_{\Delta R}$) grid with a fourth-order polynomial, we solve for
the density at which $n_{\Delta R}=n_{\tt w}.$ We find $n_{\Delta
  R}\approx10^{13.7}$ cm$^{-3}$ (indicated with an open star in Figure
\ref{fig:n}). At this density, Equation \ref{eq:dr} is satisfied, so
that the apparent change in the column density between the two
\textit{Chandra} observations is determined only by the change in
density. Under these conditions, our constraints on the ionization
parameter of the precursor MHD wind in observation 1 imply a distance
from the black hole of $R_{\Delta R}\approx10^{10.1}$ cm. This is
outside $0.1R_{\rm C},$ i.e.\ it is consistent with the location of
the thermal wind. 

\subsubsection{$\Delta R/R$ is Constant}
\label{sec:drr}
Finally, we consider an intermediate scenario in which the changes in
the column density are dictated by simultaneous changes in $n$
\textit{and} $\Delta R.$ If we suppose that $\Delta R/R$ is constant,
i.e.\ the fractional extent of the wind does not change during the
outburst, it is easy to show that the implied density $n_{\Delta R/R}$
is given by
\begin{equation}
\label{eq:drr}
n_{\tt w}=n_{\rm \Delta R/R}=n_{2}\,\left(\frac{\xi_{1}}{\xi_{2}}\right) 
\left(\frac{L_{\rm 2}}{L_{\rm 1}}\right)\left(\frac{N_{\rm H,1}}
{N_{\rm H,2}}\right)^{2}. 
\end{equation}
Again, we are primarily interested in the self-consistent solution,
where the column densities measured from the data with {\tt warmabs}
and the density input to {\sc xstar} satisfy Equation \ref{eq:drr}, so
that the fractional extent of the wind is preserved from one
\textit{Chandra} observation to the next. We employ the same iterative
approach and {\tt warmabs} models from Section \ref{sec:dr}; for each
model, we calculate $n_{\Delta R/R}$ using our measurements of $N_{\rm H},
~\xi,$ and $L$ (see Table \ref{tbl:mod} and the filled circles in
Figure \ref{fig:n}). The assumption that $\Delta R/R$ is constant
leads to smaller densities and larger radii than the assumption that
$n$ or $\Delta R$ is constant. To find the self-consistent solution,
we interpolate over our grid (filled circles in Figure \ref{fig:n}) as
before, and we find $n_{\Delta R/R}\approx10^{12.4}$ cm$^{-3}$ and
$R_{\Delta R/R}\approx10^{10.7}$ cm (filled stars in Figure
\ref{fig:n}). Once again, this location is also consistent with the
location of the thermal wind.  

To summarize the results of this section briefly, if we suppose that a
magnetically-driven wind makes a contribution to the Fe\,{\sc xxvi}
absorption line in observation 1, then photoionization models require
that this MHD wind have a lower column density than in observation
2. Considering self-consistent geometrical interpretations of this
difference, we find it probable that wind has lower density and
originates farther out in the first observation; in many cases its
location is consistent with that of the thermally-driven wind. Thus,
in this scenario, the wind found in the first observation may
actually be a hybrid MHD/thermal wind.

\section{SUMMARY AND CONCLUSIONS}
\label{sec:concl}
In this paper, we have explored the puzzling variability of the
accretion disk wind of the black hole GRO J1655--40, as seen by
\textit{Chandra} during the 2005 outburst. The first \textit{Chandra}
HETGS observation, taken in March of that year as the source was
beginning a transition out of the hard state, revealed a single
Fe\,{\sc xxvi} absorption line at 7 keV; in a much softer state twenty 
days later, \textit{Chandra} saw a forest of hundreds of lines
(\citealt{Kallman09}), indicative of one of the densest,
highest-column winds observed in an X-ray binary to date (so dense and
so highly ionized that magnetic processes must have been the primary
launching mechanism; \citealt{M06a,M08}). 

But how are we to interpret the relatively sudden appearance of so
many lines? Is it possible that state-dependent changes in the
ionizing spectrum dominate changes in the optical depth in the wind,
or is a strongly variable outflow required to account for the
variable lines? In Section \ref{sec:warmabs}, we showed that the wind
could not possibly have been the same in both \textit{Chandra}
observations, or else both spectra would have shown similar numbers of
strong absorption lines. In fact, not only is the broadband spectrum
in observation 1 incapable of over-ionizing the wind discovered by 
\citetalias{M06a}, it should actually be \textit{better} at driving
a thermal wind off the disk! 

Armed with the certainty that the wind must have changed during the
outburst, we considered in Section \ref{sec:discuss} some of the ways
that it might have evolved from one \textit{Chandra} observation to
the next. This was not meant to be an exhaustive study of every
conceivable variant of the geometry of the wind, but a focused
exploration of some of the cleanest explanations. In Section
\ref{sec:thermal}, we considered the possibility that there might be
two physically-distinct outflows in GRO J1655--40 (see
\citealt{Kallman09} for dynamical evidence for this conclusion), one
persistent Compton-heated wind and one transient MHD wind. In Section
\ref{sec:evolve}, we considered several alternate scenarios in which
both the thermal component and the MHD component were present in both
observations; we modeled the MHD component with a grid of {\sc xstar}
models to constrain its properties. In these scenarios, the absence of
so many lines during the first \textit{Chandra} observation indicates
a significantly smaller column density for the precursor MHD wind,
which could be due to changes in its radial extent, thickness, or
filling factor, its density, or both. In the latter two cases, the
location of the MHD component is consistent with the location of the
thermal component (implied by the measured ionization parameter and
Compton temperature, respectively). 

Although there is no compelling empirical evidence highlighting any of
these scenarios as preferable, there is tantalizing evidence from
contemporaneous \textit{XMM-Newton} observations
(\citealt{DiazTrigo07}) that the single-line wind detected in the
first \textit{Chandra} observation might evolve smoothly into the
rich, dense outflow discovered by \citet{M06a}, with a second, more
ionized and faster component mentioned by \citet{Kallman09}. Following
this line of reasoning, we suggest that the first \textit{Chandra}
observation might be probing a hybrid MHD/thermal wind from the outer
disk that evolves over the next  three weeks into two distinct flows
with very different properties. Future theoretical work on the
expected behavior of disk winds in outburst may shed more light on
these results, and whether thermal wind models can actually produce
the observed lines (see \citealt{Luketic10}). For the moment, it is
remarkable that such insights, even if tentative, can be gleaned from
a spectrum with only one line. But we have the benefit of a great deal
of context from other observations during the outburst, as well as
high-quality spectra from \textit{RXTE}. 

Regardless of the precise physical explanation, it is abundantly clear
that the differences between the two \textit{Chandra} spectra cannot
be explained by differences in the number of ionizing photons. 
Instead, the changes in the wind are very likely linked in a
fundamental way to the state of the accretion disk (see also
\citealt{L02,Schulz02,M06b,M08,NL09,Blum10,N11a,N12a}). It would be
particularly interesting if the changes in the magnetically-driven
wind could be tied to changes in the magnetic field configuration in
different states, but different accretion rates, radiative
efficiencies, and scale heights in the inner accretion disk
(e.g.\ \citealt{N12a,Ponti12}) may also be important. In a forthcoming
paper, we will perform a comprehensive study of the radio, infrared, 
and X-ray emission in order to explore the relationship between the
accretion state and the wind in much more detail.

%{\it Facilities:} \facility{RXTE(PCA)}
\acknowledgements We thank Mike Nowak, Ron Remillard, Julia Lee, and
Claude Canizares for helpful discussions, and the referee for comments
that improved the clarity of our work. J.N.\ gratefully acknowledges
funding support from the National Aeronautics and Space Administration
through the Smithsonian Astrophysical Observatory contract SV3-73016
to MIT for support of the \textit{Chandra} X-ray Center, which is
operated by  the Smithsonian Astrophysical Observatory for and on
behalf of the National Aeronautics Space Administration under contract
NAS8-03060. This research has made use of data obtained from the High
Energy Astrophysics Science Archive Research Center (HEASARC),
provided by NASA's Goddard Space Flight Center.

\bibliographystyle{apj}
\bibliography{ms}

\begin{thebibliography}{45}
\expandafter\ifx\csname natexlab\endcsname\relax\def\natexlab#1{#1}\fi

\bibitem[{{Begelman} {et~al.}(1983){Begelman}, {McKee}, \& {Shields}}]{B83}
{Begelman}, M.~C., {McKee}, C.~F., \& {Shields}, G.~A. 1983, \apj, 271, 70

\bibitem[{{Blum} {et~al.}(2010){Blum}, {Miller}, {Cackett}, {Yamaoka},
  {Takahashi}, {Raymond}, {Reynolds}, \& {Fabian}}]{Blum10}
{Blum}, J.~L., {Miller}, J.~M., {Cackett}, E., {et~al.} 2010, \apj, 713, 1244

\bibitem[{{Brandt} \& {Schulz}(2000)}]{BS00}
{Brandt}, W.~N. \& {Schulz}, N.~S. 2000, \apjl, 544, L123

\bibitem[{{Brocksopp} {et~al.}(2006){Brocksopp}, {McGowan}, {Krimm}, {Godet},
  {Roming}, {Mason}, {Gehrels}, {Still}, {Page}, {Moretti}, {Shrader},
  {Campana}, \& {Kennea}}]{Brocksopp06}
{Brocksopp}, C., {McGowan}, K.~E., {Krimm}, H., {et~al.} 2006, \mnras, 365,
  1203

\bibitem[{{Canizares} {et~al.}(2005){Canizares}, {Davis}, {Dewey}, {Flanagan},
  {Galton}, {Huenemoerder}, {Ishibashi}, {Markert}, {Marshall}, {McGuirk},
  {Schattenburg}, {Schulz}, {Smith}, \& {Wise}}]{C05}
{Canizares}, C.~R., {Davis}, J.~E., {Dewey}, D., {et~al.} 2005, \pasp, 117,
  1144

\bibitem[{{D{\'{\i}}az Trigo} {et~al.}(2007){D{\'{\i}}az Trigo}, {Parmar},
  {Miller}, {Kuulkers}, \& {Caballero-Garc{\'{\i}}a}}]{DiazTrigo07}
{D{\'{\i}}az Trigo}, M., {Parmar}, A.~N., {Miller}, J., {Kuulkers}, E., \&
  {Caballero-Garc{\'{\i}}a}, M.~D. 2007, \aap, 462, 657

\bibitem[{{Ebisawa}(1997)}]{Ebisawa97a}
{Ebisawa}, K. 1997, in X-Ray Imaging and Spectroscopy of Cosmic Hot Plasmas,
  ed. {F.~Makino \& K.~Mitsuda}, 427--+

\bibitem[{{Fender} {et~al.}(2004){Fender}, {Belloni}, \& {Gallo}}]{FBG04}
{Fender}, R.~P., {Belloni}, T.~M., \& {Gallo}, E. 2004, \mnras, 355, 1105

\bibitem[{{Houck}(2002)}]{Houck02}
{Houck}, J.~C. 2002, in High Resolution X-ray Spectroscopy with XMM-Newton and
  Chandra, ed. {G.~Branduardi-Raymont}

\bibitem[{{Houck} \& {Denicola}(2000)}]{HD00}
{Houck}, J.~C. \& {Denicola}, L.~A. 2000, in Astronomical Society of the
  Pacific Conference Series, Vol. 216, Astronomical Data Analysis Software and
  Systems IX, ed. {N.~Manset, C.~Veillet, \& D.~Crabtree}, 591--+

\bibitem[{{Jahoda} {et~al.}(1996){Jahoda}, {Swank}, {Giles}, {Stark},
  {Strohmayer}, {Zhang}, \& {Morgan}}]{J96}
{Jahoda}, K., {Swank}, J.~H., {Giles}, A.~B., {et~al.} 1996, in Society of
  Photo-Optical Instrumentation Engineers (SPIE) Conference Series, Vol. 2808,
  Society of Photo-Optical Instrumentation Engineers (SPIE) Conference Series,
  ed. {O.~H.~Siegmund \& M.~A.~Gummin}, 59--70

\bibitem[{{Kallman} {et~al.}(2009){Kallman}, {Bautista}, {Goriely}, {Mendoza},
  {Miller}, {Palmeri}, {Quinet}, \& {Raymond}}]{Kallman09}
{Kallman}, T.~R., {Bautista}, M.~A., {Goriely}, S., {et~al.} 2009, \apj, 701,
  865

\bibitem[{{King} {et~al.}(2012){King}, {Miller}, {Raymond}, {Fabian},
  {Reynolds}, {Kallman}, {Maitra}, {Cackett}, \& {Rupen}}]{King11}
{King}, A.~L., {Miller}, J.~M., {Raymond}, J., {et~al.} 2012, \apj, in press

\bibitem[{{Kotani} {et~al.}(2000{\natexlab{a}}){Kotani}, {Ebisawa}, {Dotani},
  {Inoue}, {Nagase}, {Tanaka}, \& {Ueda}}]{K00}
{Kotani}, T., {Ebisawa}, K., {Dotani}, T., {et~al.} 2000{\natexlab{a}}, \apj,
  539, 413

\bibitem[{{Kotani} {et~al.}(1997){Kotani}, {Kawai}, {Matsuoka}, {Dotani},
  {Inoue}, {Nagase}, {Tanaka}, {Ueda}, {Yamaoka}, {Brinkmann}, {Ebisawa},
  {Takeshima}, {White}, {Harmon}, {Robinson}, {Zhang}, {Tavani}, \&
  {Foster}}]{Kotani97a}
{Kotani}, T., {Kawai}, N., {Matsuoka}, M., {et~al.} 1997, in American Institute
  of Physics Conference Series, Vol. 410, Proceedings of the Fourth Compton
  Symposium, ed. {C.~D.~Dermer, M.~S.~Strickman, \& J.~D.~Kurfess}, 922--926

\bibitem[{{Kotani} {et~al.}(2000{\natexlab{b}}){Kotani}, {Ebisawa}, {Inoue},
  {Kawai}, {Matsuoka}, {Nagase}, {Robinson}, {Takeshima}, {Ueda}, {Yamaoka}, \&
  {Yoshida}}]{Kotani2000a}
{Kotani}, T., {Ebisawa}, K., {Inoue}, H., {et~al.} 2000{\natexlab{b}}, Advances
  in Space Research, 25, 445

\bibitem[{{Lee} {et~al.}(2002){Lee}, {Reynolds}, {Remillard}, {Schulz},
  {Blackman}, \& {Fabian}}]{L02}
{Lee}, J.~C., {Reynolds}, C.~S., {Remillard}, R., {et~al.} 2002, \apj, 567,
  1102

\bibitem[{{Luketic} {et~al.}(2010){Luketic}, {Proga}, {Kallman}, {Raymond}, \&
  {Miller}}]{Luketic10}
{Luketic}, S., {Proga}, D., {Kallman}, T.~R., {Raymond}, J.~C., \& {Miller},
  J.~M. 2010, \apj, 719, 515

\bibitem[{{Meier}(2001)}]{Meier01}
{Meier}, D.~L. 2001, \apjl, 548, L9

\bibitem[{{Miller} {et~al.}(2004){Miller}, {Raymond}, {Fabian}, {Homan},
  {Nowak}, {Wijnands}, {van der Klis}, {Belloni}, {Tomsick}, {Smith},
  {Charles}, \& {Lewin}}]{M04a}
{Miller}, J.~M., {Raymond}, J., {Fabian}, A.~C., {et~al.} 2004, \apj, 601, 450

\bibitem[{{Miller} {et~al.}(2006{\natexlab{a}}){Miller}, {Raymond}, {Fabian},
  {Steeghs}, {Homan}, {Reynolds}, {van der Klis}, \& {Wijnands}}]{M06a}
{Miller}, J.~M., {Raymond}, J., {Fabian}, A., {et~al.} 2006{\natexlab{a}},
  \nat, 441, 953

\bibitem[{{Miller} {et~al.}(2006{\natexlab{b}}){Miller}, {Raymond}, {Homan},
  {Fabian}, {Steeghs}, {Wijnands}, {Rupen}, {Charles}, {van der Klis}, \&
  {Lewin}}]{M06b}
{Miller}, J.~M., {Raymond}, J., {Homan}, J., {et~al.} 2006{\natexlab{b}}, \apj,
  646, 394

\bibitem[{{Miller} {et~al.}(2008){Miller}, {Raymond}, {Reynolds}, {Fabian},
  {Kallman}, \& {Homan}}]{M08}
{Miller}, J.~M., {Raymond}, J., {Reynolds}, C.~S., {et~al.} 2008, \apj, 680,
  1359

\bibitem[{{Miller} {et~al.}(2011){Miller}, {Maitra}, {Cackett},
  {Bhattacharyya}, \& {Strohmayer}}]{Miller11}
{Miller}, J.~M., {Maitra}, D., {Cackett}, E.~M., {Bhattacharyya}, S., \&
  {Strohmayer}, T.~E. 2011, \apjl, 731, L7+

\bibitem[{{Neilsen} \& {Lee}(2009)}]{NL09}
{Neilsen}, J. \& {Lee}, J.~C. 2009, \nat, 458, 481

\bibitem[{{Neilsen} {et~al.}(2011){Neilsen}, {Remillard}, \& {Lee}}]{N11a}
{Neilsen}, J., {Remillard}, R.~A., \& {Lee}, J.~C. 2011, \apj, 737, 69

\bibitem[{{Neilsen} {et~al.}(2012){Neilsen}, {Petschek}, \& {Lee}}]{N12a}
{Neilsen}, J., {Petschek}, A.~J., \& {Lee}, J.~C. 2012, \mnras, 2287

\bibitem[{{Orosz} \& {Bailyn}(1997)}]{Orosz97}
{Orosz}, J.~A. \& {Bailyn}, C.~D. 1997, \apj, 477, 876

\bibitem[{{Ponti} {et~al.}(2012){Ponti}, {Fender}, {Begelman}, {Dunn},
  {Neilsen}, \& {Coriat}}]{Ponti12}
{Ponti}, G., {Fender}, R., {Begelman}, M., {et~al.} 2012, \mnras, in press

\bibitem[{{Proga}(2000)}]{Proga2000}
{Proga}, D. 2000, \apj, 538, 684

\bibitem[{{Proga} \& {Kallman}(2002)}]{PK02}
{Proga}, D. \& {Kallman}, T.~R. 2002, \apj, 565, 455

\bibitem[{{Rahoui} {et~al.}(2010){Rahoui}, {Chaty}, {Rodriguez}, {Fuchs},
  {Mirabel}, \& {Pooley}}]{Rahoui10}
{Rahoui}, F., {Chaty}, S., {Rodriguez}, J., {et~al.} 2010, \apj, 715, 1191

\bibitem[{{Reynolds} \& {Miller}(2010)}]{ReynoldsM10}
{Reynolds}, M.~T. \& {Miller}, J.~M. 2010, \apj, 723, 1799

\bibitem[{{Saito} {et~al.}(2006){Saito}, {Yamaoka}, {Fukuyama}, {Miyakawa},
  {Yoshida}, \& {Homan}}]{Saito06}
{Saito}, K., {Yamaoka}, K., {Fukuyama}, M., {et~al.} 2006, in VI Microquasar
  Workshop: Microquasars and Beyond

\bibitem[{{Schulz} \& {Brandt}(2002)}]{Schulz02}
{Schulz}, N.~S. \& {Brandt}, W.~N. 2002, \apj, 572, 971

\bibitem[{{Shields} {et~al.}(1986){Shields}, {McKee}, {Lin}, \&
  {Begelman}}]{Shields86}
{Shields}, G.~A., {McKee}, C.~F., {Lin}, D.~N.~C., \& {Begelman}, M.~C. 1986,
  \apj, 306, 90

\bibitem[{{Steiner} {et~al.}(2009){Steiner}, {McClintock}, {Remillard},
  {Narayan}, \& {Gou}}]{Steiner09a}
{Steiner}, J.~F., {McClintock}, J.~E., {Remillard}, R.~A., {Narayan}, R., \&
  {Gou}, L. 2009, \apjl, 701, L83

\bibitem[{{Tarter} {et~al.}(1969){Tarter}, {Tucker}, \& {Salpeter}}]{Tarter69}
{Tarter}, C.~B., {Tucker}, W.~H., \& {Salpeter}, E.~E. 1969, \apj, 156, 943

\bibitem[{{Ueda} {et~al.}(2004){Ueda}, {Murakami}, {Yamaoka}, {Dotani}, \&
  {Ebisawa}}]{U04}
{Ueda}, Y., {Murakami}, H., {Yamaoka}, K., {Dotani}, T., \& {Ebisawa}, K. 2004,
  \apj, 609, 325

\bibitem[{{Ueda} {et~al.}(2009){Ueda}, {Yamaoka}, \& {Remillard}}]{U09}
{Ueda}, Y., {Yamaoka}, K., \& {Remillard}, R. 2009, \apj, 695, 888

\bibitem[{{Wilms} {et~al.}(2000){Wilms}, {Allen}, \& {McCray}}]{Wilms00}
{Wilms}, J., {Allen}, A., \& {McCray}, R. 2000, \apj, 542, 914

\bibitem[{{Woods} {et~al.}(1996){Woods}, {Klein}, {Castor}, {McKee}, \&
  {Bell}}]{Woods96}
{Woods}, D.~T., {Klein}, R.~I., {Castor}, J.~I., {McKee}, C.~F., \& {Bell},
  J.~B. 1996, \apj, 461, 767

\bibitem[{{Zdziarski} {et~al.}(1996){Zdziarski}, {Johnson}, \&
  {Magdziarz}}]{Zdziarski96}
{Zdziarski}, A.~A., {Johnson}, W.~N., \& {Magdziarz}, P. 1996, \mnras, 283, 193

\bibitem[{{Zimmerman} {et~al.}(2005){Zimmerman}, {Narayan}, {McClintock}, \&
  {Miller}}]{Zimmerman05}
{Zimmerman}, E.~R., {Narayan}, R., {McClintock}, J.~E., \& {Miller}, J.~M.
  2005, \apj, 618, 832

\bibitem[{{{\.Z}ycki} {et~al.}(1999){{\.Z}ycki}, {Done}, \& {Smith}}]{Zycki99}
{{\.Z}ycki}, P.~T., {Done}, C., \& {Smith}, D.~A. 1999, \mnras, 309, 561

\end{thebibliography}

\label{lastpage}

\end{document}